\documentclass[12pt]{article}
\usepackage{epsf}
\title{\bf On the numerical method of Casimir energy renormalization in 
the presence of logarithmical divergencies } \author{ Il.  Malakhov$^{a,}$ 
\thanks{e-mail address:  malakhov@goa.bog.msu.ru }\ ,  \ P. Silaev$^{a,b}$ 
\ , K.  Sveshnikov$^{a,b}$ \\ {\em $^{a}$Physics Department and 
$^{b}$Institute of Theoretical MicroPhysics,} \\  {\em Moscow State 
University, Moscow 119992, Russia} }

\begin{document}

\maketitle

\abstract{\small A non-subtractive recipe of Casimir energy 
renormalization efficient in the presence of logarithmically divergent 
terms is proposed.  It is demonstrated that it can be applied even then, 
when energy levels can be obtained only numerically and neither their 
asymptotical behavior, nor the analytic form of spectral equation is 
known. The results of calculations performed with this method are compared 
to those obtained by means of explicit subtraction of divergent terms from 
energy.}

\normalsize

\section{Introduction}
   
Ever since Casimir [1] has obtained corrections to the energy of a
macroscopic system due to vacuum fluctuations of quantized electromagnetic 
field in 1948 this effect has been intensively studied both from 
theoretical and experimental points of view. Nevertheless the calculation 
of Casimir energy except for the most simple problems involving free 
fields inside cavities with flat boundaries is quite non-trivial yet. To 
realize this, recall a great number of papers devoted to a free field in 
the interior of a sphere [2-11]. Despite the fact that in this case one can 
explicitly write out spectral equations, the first analytical results have 
been obtained only in [12] for massive scalar field and in [13] for 
fermions.

It should be stressed that the knowledge of analytical form of spectral 
equation has been crucial in [12] since it makes possible the transition 
from the sums containing the unknown energy levels to the integrals with 
the explicit integrands [14]. The goal of this paper is to demonstrate a 
method which can be applied to numerical calculation of Casimir energy in 
cases when these requirements are not met. Moreover, we are not going to 
use a rather standard trick [7], [15], which lets one overcome problems 
arising due to the presence of logarithmic divergency in Casimir energy of 
free massless fields inside spherical shells. Note that logarithmic 
divergency appears as a consequence of a curved surface bounding the shell 
and makes the energy renormalization ambiguous. The main idea of the trick 
is to consider the "inner" and "extra" problems together since their 
logarithmic divergencies cancel each other.

The ambiguity of Casimir energy renormalization in the presence of 
logarithmic divergency is quite obvious. Indeed, in case of massless 
fields the energy of the system can be characterized by a single 
dimensional parameter $L$ which is the linear size of the system. The 
regularization parameter $\alpha$ can be also chosen to have a dimension 
of length. In the absence of logarithmic divergency the "minimal 
subtraction" of singular terms is not only natural but also well grounded. 
Indeed, any term proportional to $\alpha^{-s}$ ($s>0$) is obviously 
proportional to $L^{s-1}$, i.e. to the non-negative power of $L$. This 
makes it possible to normalize the final result at $L=\infty$, where 
Casimir energy should become zero, and subtract all singular terms at the 
same time.  After such subtraction the only remaining term in the limit 
$\alpha\to 0$, which reads $c\alpha^0/L$, provides the final result.

In the presence of logarithmic divergency the subtraction becomes 
ambiguous, since in order to renormalize the term $c \alpha^0 
\log(\alpha/L)/L$, one should subtract $c \alpha^0 \log(d \alpha 
/L)/L$, where $d$ is an arbitrary constant, which cannot be determined 
from the normalizing condition at $L\to \infty$. 

\section{Massive scalar field in 1D}

To illustrate the main idea of the proposed recipe let's consider 
Casimir energy with the logarithmic divergency in the most simple case, 
i.e. Casimir energy of massive scalar field on interval of length $L$ with 
Dirichlet boudary conditions at the ends of the interval. It reads:
$$ {\cal E}_{cas}={1\over 2}\sum_{n=1}^\infty  \omega_n ={1\over 2} 
\sum_{n=1}^\infty  \sqrt{(\pi n/L)^2+m^2} \eqno (1)   $$

\noindent Note that we aren't going to subtract the Minkowski vacuum 
contribution from $(1)$ as proposed in "standard" approaches. Instead we 
will directly pass to renormalizaton.

The regularization of $(1)$ requires the introduction of the parameter 
$\alpha$ which has a dimension of length and stands in the argument 
of the cut-off function $F(\alpha \omega_n)$:

$$ {\cal E}_{cas}^{(r)}={1\over 2}\sum_{n=1}^\infty  \omega_n F (\alpha 
\omega_n) \eqno (2)$$

\noindent Trivial considerations based on dimensional analysis lead to the 
following expression for the regularized Casimir energy $${\cal 
E}_{cas}^{(r)}\simeq c_{-2}{ L \over \alpha^2} + c_{-1}{ L^0 \over 
\alpha^1} + c_{0}{ 1 \over L} + c_{\lambda}{ m^2 L \log(\alpha/L) } + 
\cdots \eqno (3)$$

\noindent It can be easily verified that for various cut-off functions 
such as $F(x)=\exp(-x)$, $F(x)=\exp(-x^2)$, $F(x)=\exp(-x^3)$, $\ldots$, 
$F(x)=\exp(-x^6)$, $F(x)=\exp(-2\cosh(x)+2)$, $\ldots$, identical 
$c_{\lambda}$ are obtained, while $c_0$ are different. The identity of 
$c_{\lambda}$ for different $F(x)$ can be demonstrated with the following 
estimation for the sum giving rise to the logarithmic divergency:  $$ 
{1\over 2}\sum_{n=1}^\infty {1\over 2} {m^2\over \omega_n } F (\alpha 
\omega_n) \sim  {m^2 L\over 4\pi } \log N \sim {m^2 L\over 4\pi } \log 
(\Delta x L/\alpha) =$$ $${m^2 L\over 4\pi } \log (L/\alpha)  + {m^2 
L\over 4\pi } \log (\Delta x) \; ,  \eqno (4)$$ where $N\sim \Delta x 
L/\alpha$ and $\Delta x$ is a cut-off interval of $F(x)$.

Since any subtraction in the presence of logarithmic divergency is 
ambiguous this procedure should be excluded from consideration along with 
the logarithmic divergency itself. To realize that let's calculate 
$\partial_L^2 {\cal E}_{cas}^{(r)}$:

$$\partial_L^2 {\cal E}_{cas}^{(r)}\simeq c_{0}{ 2 \over L^3} + 
c_{\lambda}{ m^2 \over L } + \cdots \eqno (5) $$

\noindent The obtained expression is regular in the limit $\alpha \to 0$, 
so no subtraction is required. The knowledge of the function $\partial_L^2 
{\cal E}_{cas}^{(r)}$, lets one reconstruct the required ${\cal 
E}_{cas}^{(r)}$ unambiguously, since the initial conditions at 
$L\to\infty$ are well-known:  both ${\cal E}_{cas}^{(r)}$ and $\partial_L 
{\cal E}_{cas}^{(r)}$ should become zero.  Note that while $(5)$ doesn't 
describe the asymptotical behavior of $\partial_L^2 {\cal E}_{cas}^{(r)}$ 
at $L\to\infty$, it demonstrates the disappearence of all singular terms 
in it. Moreover the following integral approximation shows that 
$\partial_L^2 {\cal E}_{cas}^{(r)}$ vanishes for $L \to \infty$: $$ {\cal 
E}_{cas}^{(r)}={1\over 2}\sum_{n=1}^\infty \omega_n F (\alpha \omega_n) = 
{1\over 4}\sum_{n=-\infty}^\infty  \omega_n F (\alpha \omega_n) -{m\over 
4} F(\alpha m) \approx $$  $$ \approx {1\over 2} \int_{-\infty}^\infty dx 
(L/\pi) \sqrt{x^2+m^2} F (\alpha \sqrt{x^2+m^2}) -{m\over 4} F(\alpha m) 
\eqno(6) $$

\section{Method of calculation in general case}

Let's generalize the proposed method in such a way that it doesn't require 
the analytical expression for energy levels.  Suppose one has a set of 
energy levels  of some spectrum $\omega_n$ and the corresponding Casimir 
energy contains the logarithmic divergency.  First of all it turns out to 
be possible to modify the initial expression for the Casimir energy by 
introduction of some parameter $\mu$ in such a way that $$ {1\over 
2}\sum_{n=1}^\infty  \sqrt{\omega_n^2-\mu^2} F (\alpha 
\sqrt{\omega_n^2-\mu^2}) \eqno (7)$$
doesn't contain the logarithmic divergency. For the massive 
scalar field on an interval $\mu$ is obviously equal to the mass of the 
field.  In less trivial three-dimensional cases with spherical symmetry 
$\mu$ is some parameter having a dimension of mass which characterizes the 
total coefficient by the logarithmic divergency with all values of angular 
momentum taken into account.

The next step is to introduce an "additional mass" of the field ${\cal M}$ 
and study the Casimir as a function of it in the range from ${\cal M}=0$ 
to ${\cal M}=\infty$. In fact it's helpful to introduce another parameter 
$M$:  $M^2 \equiv {\cal M}^2+\mu^2$ and study the modified Casimir energy

$$ {\cal E}_{cas} (M) = {1\over 2}\sum_{n=1}^\infty \sqrt{\omega_n^2+ 
M^2-\mu^2} \eqno(8)$$

\noindent as a function of $M$ in the range from $M=\infty$ to $M=\mu$. To 
realize this one should calculate numerically the following quantity in 
the specified range of $M$:

$$ \partial^2_M  \left( {\cal E}_{cas}^{(r)} ({M})/M \right) =$$

$$ = \partial^2_M  \left[ {1\over 2}\sum_{n=1}^\infty  \sqrt{ 
{\omega_n^2-\mu^2 \over M^2} + 1\; } \;\; F \left(\alpha \sqrt{ 
{\omega_n^2-\mu^2 \over M^2} +1\;} \, \right) \right] \eqno(9) $$

\noindent Note that in contrast to our previous considerations we have 
substituted dimensionless quantity

$$ {1 \over M}  \sqrt{ \omega_n^2 + M^2-\mu^2   \; }= \sqrt{ 
{\omega_n^2-\mu^2 \over M^2} + 1\; } $$

\noindent to the argument of $F(x)$, so that $\alpha$ should be also taken 
dimensionless.

An alternative interpretation  of $(9)$ follows from the observation that 
$M$ acts as an effective length $L$ in the expression for Casimir energy. 
Indeed, $(9)$ can be obtained from the initial expression as a result 
of the folowing transformation of the spectrum. At the first step 
$\omega_n$ is transformed to $\omega_n' = \sqrt{ \omega_n^2-\mu^2 }/\mu$, 
which is dimensionless spectrum with the subtracted effective mass.  After 
that the scale transformation of the system $x\to x (M/\mu)$ 
and $\omega_n'\to \omega_n'' =\omega_n' /(M/\mu)$ is performed. In the end 
the unit mass is "added" to the obtained spectrum:

$$\omega_n''\to \sqrt{ (\omega_n')^2+1}= \sqrt{ {\omega_n^2-\mu^2 \over 
M^2} + 1\; } \eqno(10)$$

\noindent The limit $M\to\infty$ obviously corresponds to the infinite 
size of the system, while for $M=\mu$ one obtains the initial spectrum 
divided by $\mu$.
                                                        
It's easy to see that all divergent terms in $(9)$ vanish. In the 
limit $n\to \infty$ one can make use of the following expansion $$ \sqrt{ 
{\omega_n^2-\mu^2 \over M^2} + 1\; }\approx { \sqrt{ \omega_n^2-\mu^2} 
\over M } + { M \over 2 \sqrt{ \omega_n^2-\mu^2} } + \cdots \eqno(11)$$ 

\noindent The first term in this expansion gives rise to the sum which is 
free of logarithmic divergency due to the definition of $\mu$. Other 
divergencies are proportional to $(M/\alpha)^2/M$ and $(M/\alpha)^1/M$ 
and vanish when the second-order derivative is calculated.  The second 
term leads to the logarithmic divergency with the coefficient proportional 
to $M^1$ by it, which also vanishes.

As a result one has $(9)$ regular for $\alpha\to 0$ and the natural 
normalizing condition ${\cal E}_{cas} ( M \to \infty)=0$.  The latter can 
be understood from two different points of view. On one hand the quantized 
field with the infinitely large mass should have zero Casimir energy. On 
the other hand the Casimir energy in the limit of the infinite size of the 
system should become zero. Whichever interpretation is chosen, the 
obtained results let one reconstruct the required ${\cal E}_{cas} 
(M=\mu)$ which corresponds to the initial spectrum.

Note, that principally one could consider ${\cal E}_{cas}^{(r)} ({M})$ 
instead of ${\cal E}_{cas}^{(r)} ({M})/M$. However that would increase the 
order of derivative  required to exclude all divergent terms by one what 
is undesirable from the practical point of view.

The proposed method turns out to be efficient not only in the 
most trivial one-dimensional cases but also in more realistic 
three-dimensional ones. However to employ it in three-dimensional case one 
should inevitably calculate the fourth-order derivative of the Casimir 
energy ($9$) since the main singular term, which is proportional to 
the volume of the system, reads $ c_{-4} L^3/\alpha^4 $. It should be also 
noted that in this case the calculations of the sums become more 
sophisticated since the final value of a typical sum is about 40 orders 
lower than intermediate values obtained during its calculation and extra 
floating-point precision is required. 

\vskip 5 true mm

\section{Numerical results}

\vskip 2 true mm
  
For scalar field on an interval $[0;L]$ with $L=1$ spectrum reads
$$\omega_n = \sqrt{ { (\pi n)^2 \over L^2 } + m^2 } \eqno (12)$$

The result of the straightforward application of our method with various 
cut-off functions such as $F(x)=\exp(-x)$, $F(x)=\exp(-x^2)$, 
$F(x)=\exp(-x^3)$, $\ldots$, $F(x)=\exp(-x^6)$, $F(x)=\exp(-2\cosh(x)+2)$ 
is presented on Fig.1. It has been shown that for each of these functions 
the same result is obtained and what's more the precision of coincidence 
depends only on the number of energy levels taken into account and the 
number of right digits used in the realization of floating-point 
arithmetics as well.

As to dependence of the Casimir energy on the mass of the field some 
important aspects should be stressed. First of all in the limit $m \to 0$ 
a well-known result for the massless scalar field is obtained.  In the 
range of large values of $m$ Casimir energy decreases exponentially as 
$e^{-2mL}$ what could be expected from qualitative considerations. The 
results obtained with our method in this trivial case are in agreement 
with those obtained using the traditional subtractive technique.

To demonstrate how the method can be employed in less trivial cases we 
consider the massless scalar field inside of a spherical shell of 
radius $R=1$. In this case the same set of cut-off functions has been 
used. The values of an effective mass $\mu=0.1377$ obtained with each of 
these functions coincide up to the first four digits. Consequently the 
precision of the obtained $E_{cas}=3.790\cdot 10^{-3}$ has the same order, 
what corresponds to about 200 $s$-levels taken into account during the 
calculations. The number of energy levels taken into account is directly 
affected by the range which the regularization parameter $\alpha$ used in 
the calculations belongs to. Therefore one can control precision of the 
final result simply changing the range of employed values of the 
regularization parameter.

Note that in the framework of this approach we have obtained not only 
Casimir energy of the massless scalar field (corresponding to ${\cal 
M}=0$) but also Casimir energy for all possible values of mass in the 
range from zero to the "effective" infinity. The dependence of the Casimir 
energy of the scalar field inside the sphere on the mass of the field is 
presented on Fig.2.

It should be stressed that on the contrary to the results obtained with 
the traditional subtractive technique in [12] our result doesn't contain
logarithmical singularity at ${\cal M}=0$ what seems more reasonable from 
physical point of view. The most likely explanation of this difference is 
that the subtractive procedure contains some arbitrariness. As a result 
some function which has regular behavior at ${\cal M} \to \infty$ but is 
singular at ${\cal M} \to 0$ could be subtracted from the final result.

As has been pointed out in [12,15] there is no argument at present which 
can remove this arbitrariness in case of a massless scalar field inside of 
a sphere. Therefore Casimir effect in the whole space with Dirichlet 
boundary conditions on the sphere is usually considered. It seems 
reasonable to calculate the Casimir energy in the last case employing our 
method.

In fact there are two ways to proceed to take exterior into account. The 
first one deals with the continuous spectrum and requires that all the 
regularized sums be replaced with the appropriate integrals containing the 
energy levels density in the integrand. The second way lets one work with 
discrete spectrum all the time.  To realize that one should place the 
initial spherical shell into another sphere with the radius 
$R_{out}=kR_{in}$ where $k \ge 1$ and calculate the Casimir energy for 
the system bounded by the outer sphere taking into account boundary 
conditions on both of the spheres.  For each finite $k$ the spectrum is 
discrete and the developed technique can be applied without modification. 
The required result can be achieved in the limit $k \to \infty$. In 
practice it turns out that for $k \geq k_0$, where $k_0$ is finite and 
depends on the required precision, the result doesn't depend on $k$. It 
turned out that in the considered case in order to get 4 right digits in 
the final result $k_0 \simeq 10$ is quite enough.

The final result of the calculations is presented on Fig.3. Note that 
while the qualitative behavior of Casimir energy is the same as that 
obtained with methods employing explicit subtraction [15], there is no 
absolute coincidence. For example, for the massless field the result 
obtained with our recipe is ${\cal E}_{cas}({M=0}) = 0.0039$ while 
direct subtraction leads to ${\cal E}_{cas}({M=0}) = 0.0028$.

\section{Conclusion}
To summarize, an efficient technique for numerical calculation of Casimir 
energy in the presence of logarithmical divergencies has been developed. 
The advantages of the proposed method are its ideological triviality and 
universality which let one apply it to a wide range of problems in which 
numerical values for energy levels can be obtained. The results of its
application to a number of problems appear to be reasonable, especially in 
case of a curved boundary. As to disadvantages they are purely technical:  
one should employ floating-point arithmetics with extra precision to carry 
out calculations in realistic cases.

\section{Acknowledgments}
This work has been supported in part by the RF President Grant 1450.2003.2. 
One of us (Il.M.) is indebted to the Organizers of QUARKS'2004 for hospitality 
and financial support. We are also grateful to O.Pavlovski and I.Cherednikov 
for fruitful discussions.

\eject
\begin{figure}
{
\epsfbox{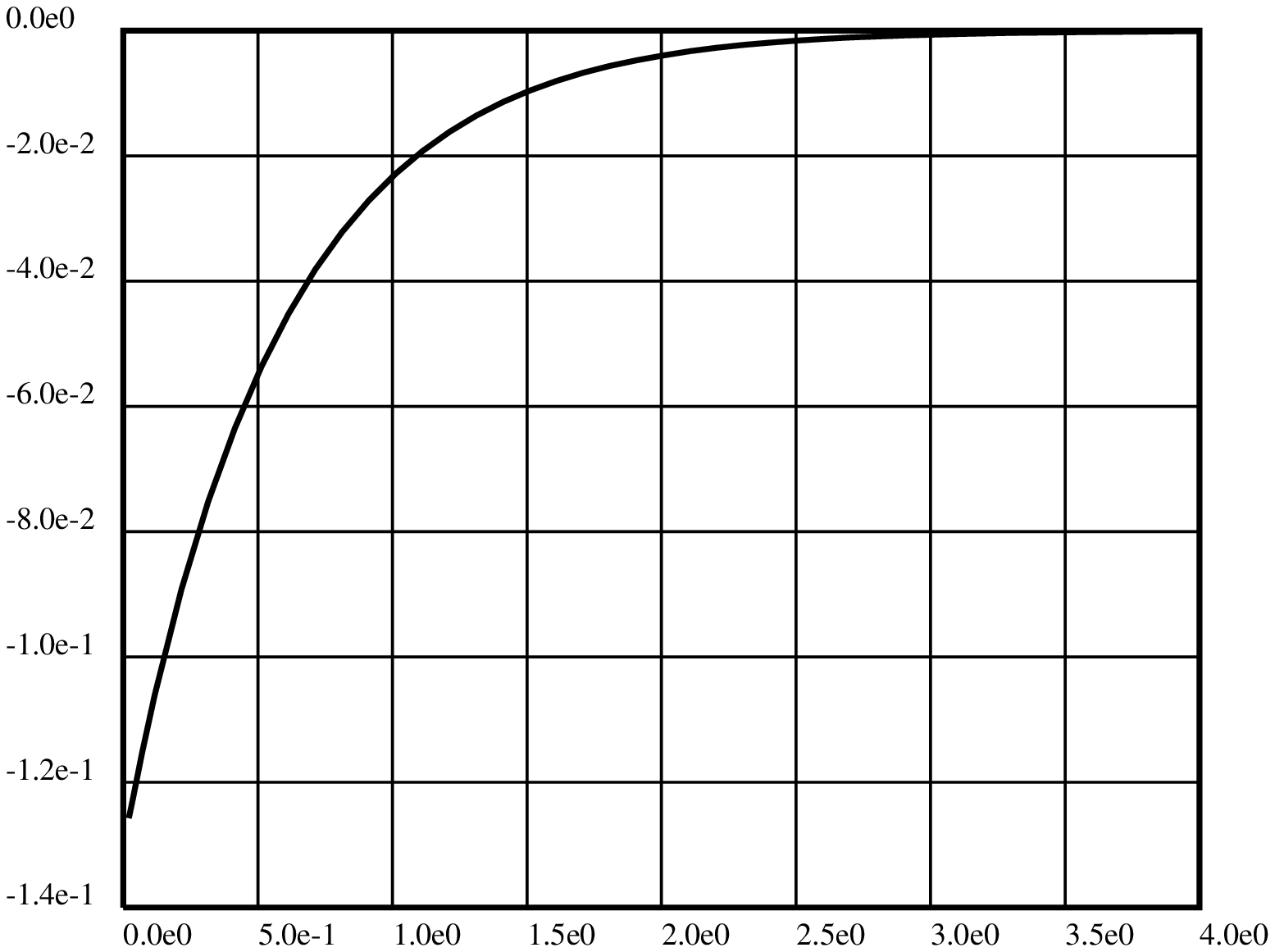}
}
Fig.1. The Casimir energy of the massive scalar field on a unit interval 
as a function of the mass of the field.
\end{figure}

\begin{figure}
{
\leftskip -15.5cm
\epsfbox{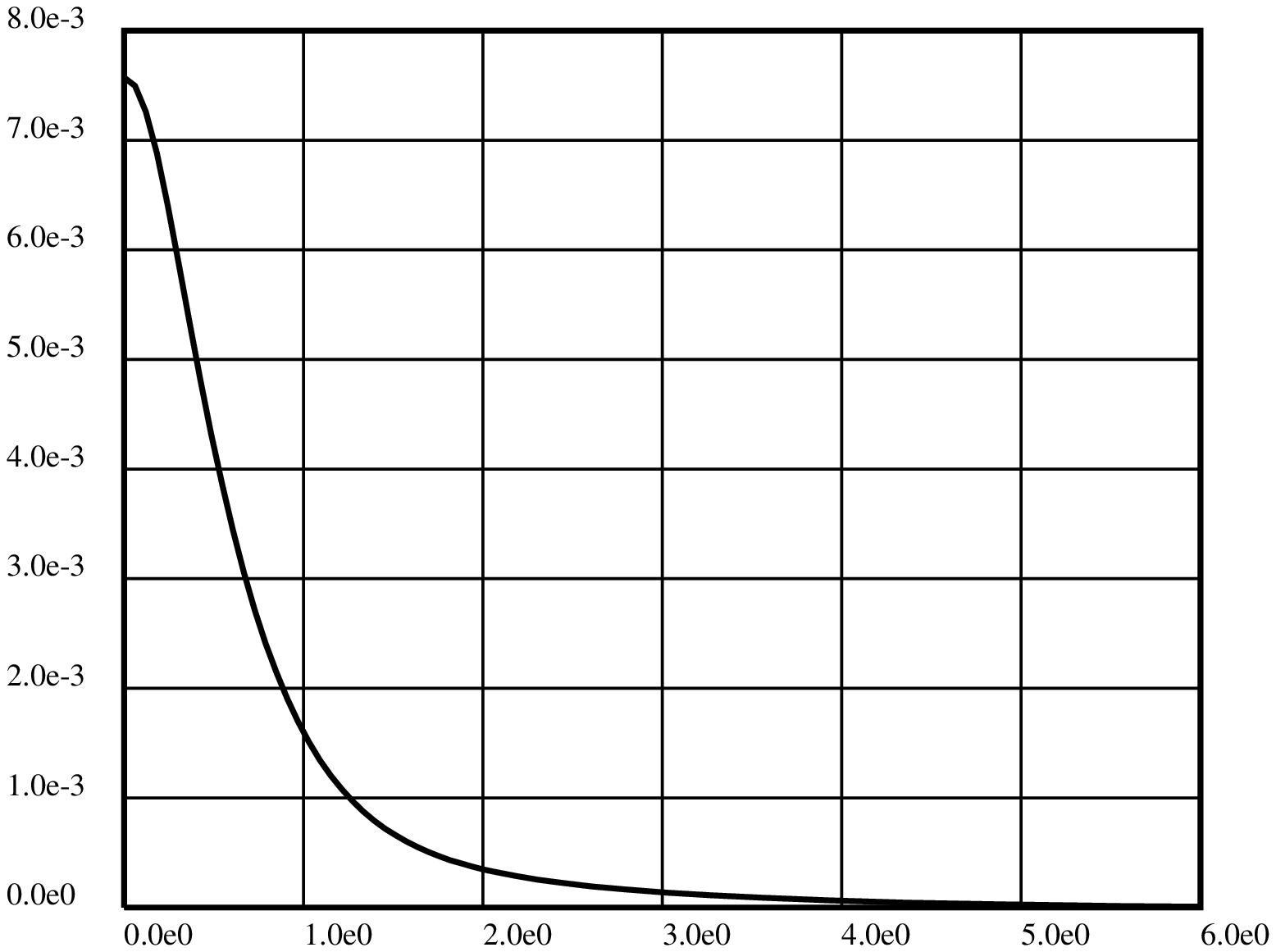}
}
Fig.2. The Casimir energy of the scalar field inside of a spherical shell 
of radius 1 obeying Dirichlet boundary conditions as a function of the 
mass of the field.
\end{figure}

\begin{figure}
{
\leftskip -1.5cm
\epsfbox{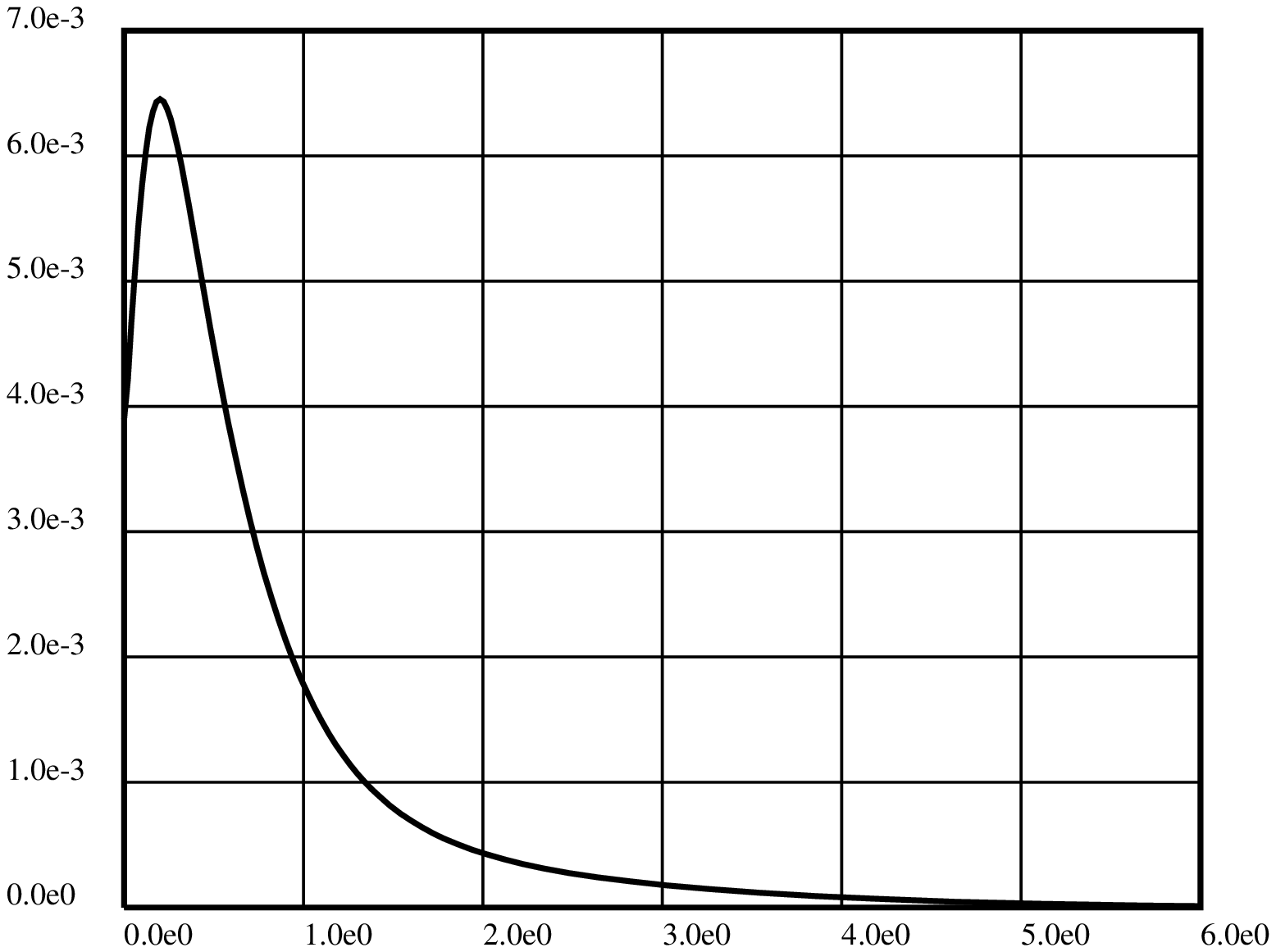}
Fig.3. The Casimir energy of a massive scalar field in the whole space 
with Dirichlet boundary condition on a sphere of radius 1 as a function of 
the mass of the field.

}
\end{figure}

\end{document}